\documentclass[aps,prb,twocolumn,floatfix,superscriptaddress,amsfonts,amssymb,amsmath,preprintnumbers]{revtex4}
\usepackage{bm}
\usepackage{psfig}
\usepackage{color}

\newcommand{\secref}[1]{Sec.~\ref{#1}}

\newcommand{\Eq}[1]{Equation~(\ref{#1})}
\newcommand{\eq}[1]{Eq.~(\ref{#1})}

\newcommand{\eqsdash}[2]{Eqs.~(\ref{#1}--\ref{#2})}

\newcommand{\Figref}[1]{Fig.~\ref{#1}}
\newcommand{\figref}[1]{Fig.~\ref{#1}}

\newcommand{\tabref}[1]{Table~\ref{#1}}

\newcommand{\bea}{\begin{eqnarray}}
\newcommand{\eea}{\end{eqnarray}}
\newcommand{\bal}{\begin{aligned}}
\newcommand{\eal}{\end{aligned}}
\newcommand{\bga}{\begin{gathered}}
\newcommand{\ega}{\end{gathered}}

\newcommand{\lt}{\left}
\newcommand{\rt}{\right}
\newcommand{\bl}{\bigl}
\newcommand{\br}{\bigr}
\newcommand{\la}{\langle}
\newcommand{\ra}{\rangle}

\newcommand{\const}{\text{const}}

\newcommand{\dd}{\partial}
\newcommand{\vdel}{\boldsymbol{\nabla}}

\newcommand{\vu}{{\bf u}}
\newcommand{\dvu}{\delta\vu}
\newcommand{\dvuperp}{\dvu_\perp}

\newcommand{\vB}{{\bf B}}
\newcommand{\vK}{{\bf K}}
\newcommand{\vb}{{\bf\hat b}}

\newcommand{\vk}{{\bf k}}

\newcommand{\kperp}{k_\perp}
\newcommand{\kpar}{k_\parallel}

\newcommand{\vperp}{v_\perp}

\newcommand{\pperp}{p_\perp}
\newcommand{\ppar}{p_\parallel}

\renewcommand{\Re}{\text{Re}}
\newcommand{\Rm}{\text{Rm}}
\newcommand{\Reeff}{\text{Re}_\text{eff}}

\newcommand{\Bsq}{\la B^2\ra}
\newcommand{\usq}{\la u^2\ra}

\newcommand{\nui}{\nu_{ii}}
\newcommand{\nusc}{\nu_\text{scatter}}
\newcommand{\nueff}{\nu_\text{eff}}
\newcommand{\gmax}{\gamma_\text{max}}
\newcommand{\nuicm}{\mu_\text{ICM}}
\newcommand{\nupar}{\mu_\parallel}
\newcommand{\nupareff}{\mu_{\parallel,\text{eff}}}
\newcommand{\mfp}{\lambda_\text{mfp}}
\newcommand{\mfpeff}{\lambda_\text{mfp,eff}}
\newcommand{\vth}{v_{\text{th},i}}
\newcommand{\Bseed}{B_\text{seed}}
\newcommand{\Bstr}{B_\text{straight}}
\newcommand{\Bcorner}{B_\text{corner}}
\newcommand{\Beq}{B_\text{eq}}
\newcommand{\Bvisc}{B_\text{visc}}
\newcommand{\rhoeq}{\rho_{i,\text{eq}}}
\newcommand{\Omeq}{\Omega_{i,\text{eq}}}
\newcommand{\betaeq}{\beta_\text{eq}}
\newcommand{\Brms}{B_\text{rms}}
\newcommand{\lB}{l_B}
\newcommand{\lpar}{l_\parallel}
\newcommand{\lperp}{l_\perp}
\newcommand{\lres}{l_\text{res}}
\newcommand{\lvisc}{l_\text{visc}}
\newcommand{\tvisc}{t_\text{visc}}

\begin{document}

\preprint{Invited talk, 47th APS DPP Meeting, Denver, Oct.~24--28, 2005; 
{\em Phys.\ Plasmas}~{\bf 13}, 056501 (2006) [{\tt astro-ph/0601246}]}

\title{Turbulence, magnetic fields and plasma physics in clusters of galaxies}
\author{A.\ A.\ Schekochihin}
\email{as629@damtp.cam.ac.uk}
\affiliation{DAMTP, University of Cambridge, Cambridge CB3 0WA, UK}
%\affiliation{King's College, Cambridge CB2 1ST, UK}
\author{S.\ C.\ Cowley}
\affiliation{Department of Physics and Astronomy, 
UCLA, Los Angeles, California 90095-1547}
\affiliation{Plasma Physics Group, Imperial College, 
Blackett Laboratory, Prince Consort Road, London~SW7~2BW, UK}
\date{\today}

\begin{abstract}
Observations of galaxy clusters show that 
the intracluster medium (ICM) is likely to be turbulent and is certainly magnetized. 
The properties of this magnetized turbulence are determined both by fundamental nonlinear 
magnetohydrodynamic 
interactions and by the plasma physics of the ICM, which has very low collisionality. 
Cluster plasma threaded by weak magnetic 
fields is subject to firehose and mirror instabilities. These saturate and produce  
fluctuations at the ion gyroscale, which can scatter particles, increasing   
the effective collision rate and, therefore, the effective Reynolds number of the ICM. 
A simple way to model this effect is proposed. The model yields 
a self-accelerating fluctuation dynamo whereby the field grows explosively fast, 
reaching the observed, dynamically important, field strength in a fraction of 
the cluster lifetime independent of the exact strength of the seed field. 
It is suggested that the saturated state of the cluster turbulence 
is a combination of the conventional isotropic magnetohydrodynamic turbulence, 
characterized by folded, direction-reversing magnetic fields 
and an Alfv\'en-wave cascade at collisionless scales. 
An argument is proposed to constrain the reversal scale of the folded field. 
The picture that emerges appears to be in qualitative agreement with 
observations of magnetic fields in clusters.  
\end{abstract}

%\pacs{52.30.Gz,52.35.Ra,95.30.Qd}

\maketitle

\section{Introduction}

Clusters of galaxies are vast and varied objects that have long attracted 
the attention of both observers (in recent decades spectacularly 
aided by X-ray and radio telescopes) and theoreticians. 
The observed properties of clusters have proved far from easy to explain as new 
data has confounded many old theories. 

The overall budget of the cluster constituents is roughly as follows: $\sim75\%$ of cluster mass 
is dark matter, whose sole function is assumed to be to provide the gravitational well, 
$\sim20\%$ of cluster mass is the diffuse X-ray-emitting plasma (the intracluster 
medium, or ICM), while the galaxies have an all but negligible mass. 
The plasma that makes up the ICM is made of hot and tenuous ionized 
hydrogen: temperatures are in the range of $1-10$~keV, number densities 
$\sim 10^{-1}-10^{-3}~\text{cm}^{-3}$, with colder, denser material found in 
the cool cores and hotter, more diffuse one in the outlying regions.\footnote{The cool 
cores occupy roughly the central 100~kpc, compared with the overall cluster 
size of order $1$~Mpc. Such a structure is, in fact, characteristic of 
the older type of clusters known historically as cooling-flow clusters 
(the more observationally popular members of this group are Hydra~A, Centaurus and Perseus), 
while younger, less dynamically relaxed clusters (e.g., Coma) have flatter 
density and temperature profies. A good summary of the parameters in a number of 
cooling-flow clusters and the relevant references can be found in 
Ref.~\onlinecite{Ensslin_Vogt_cores}.} 

Necessarily, the first observations and the first 
theoretical models of clusters concerned what one might call large-scale features, 
such as the overall profiles of mass and temperature, 
the structure formation and the role of the central objects. 
As observations increased in accuracy 
and resolution, the ICM was revealed to be much richer than simply a dull 
cloud of X-ray glow smoothly petering out with distance from the center. 
A great panoply of features has been detected: bubbles, filaments, ripples, 
edges, shocks, sound waves, etc., as well as very chaotic density, temperature and 
abundance distributions.\cite{Fabian_etal_Perseus1,Fabian_etal_Perseus3,Fabian_etal_Centaurus,Schuecker_etal,Churazov_etal_Perseus1} It is particularly the presence of chaotic fields and the evidence that 
this chaos exists in a range of scales\cite{Schuecker_etal,Vogt_Ensslin1,Vogt_Ensslin2} 
that makes one expect that ICM, like so many other astrophysical plasmas, is in a 
turbulent state. It is essential to know the properties of this turbulence in order 
to predict the current and future statistical measurements of the plasma  
and magnetic fields (spectra, correlation and distribution functions, etc.) 
and to model correctly the transport processes in the ICM that determine, for 
example, the overall temperature profiles.\cite{Voigt_Fabian,Dennis_Chandran} 

We shall assume that the physics of small scales is, 
at least to some degree, independent of large-scale circumstances 
and that we can, therefore, gain some useful understanding of the turbulence 
in clusters by ignoring large-scale features and considering 
a homogeneous subvolume of the ICM. 
In what follows, after reviewing briefly what is known 
about fluid motions (\secref{sec_turb}) and magnetic fields (\secref{sec_mag}) 
in clusters, we describe, mostly 
in qualitative terms, what we consider to be the essential aspects 
of the small-scale physics of the ICM (for plasma physics, see 
\secref{sec_plasma}). This will lead us to a tentative 
overall picture of the structure of the ICM turbulence (\secref{sec_sat}) 
as well as of the origin of its magnetic component (\secref{sec_dynamo}). 
We must emphasize that 
the current state of the debates on the nature of turbulence in clusters 
(and, indeed, on its very existence\cite{Fabian_etal_Perseus2}) is such 
that even a fundamental, conceptual view of the problem has not yet 
been agreed, and, therefore, a quantitative theory --- even an incomplete 
one such as exists for hydrodynamic turbulence --- remains 
a matter of future work. 
%In plainer words, we are very far from arguing about 
%scaling exponents of high-order structure functions as we still do not even know 
%what exactly the energy-containing and dissipative scales are or how 
%they are determined. 

\section{Turbulence}
\label{sec_turb}

The failure of one of the instruments on the ASTRO-E2 satellite has set off 
the planned direct detection of cluster turbulence\cite{Sunyaev_Norman_Bryan,Inogamov_Sunyaev} 
into the (probably not very distant) future. However, indirect evidence 
of turbulent gas motions does exist: three recent examples are the broad spectrum of pressure 
fluctuations measured in the Coma cluster\cite{Schuecker_etal}, 
detection of subsonic gas motions in the core of the Perseus 
cluster,\cite{Churazov_etal_Perseus2} and, again for the Perseus cluster,  
the broadening (assumed to be caused by turbulent diffusion) 
of abundance peaks associated with 
the brightest cluster galaxies.\cite{Rebusco_etal} 
These and other studies and models based on 
observational data appear to converge in expecting 
turbulent flows with rms velocities in the range 
$U\sim10^2-10^3$~km/s at the outer scales $L\sim10^2$~kpc. 
The energy sources for this turbulence are probably the cluster and subcluster 
merger events and/or, especially 
for the turbulence in cool cores, the active galactic nuclei (AGN). 
The aforementioned observational estimates of the strength and scale of the turbulence 
are in order-of-magnitude agreement with the outcomes of numerical simulations 
of cluster formation\cite{Norman_Bryan,Sunyaev_Norman_Bryan,Ricker_Sarazin} 
and of the buoyant rise of radio bubbles generated by the AGN.\cite{Churazov_etal_bubbles,Fujita} 
Further discussion and references on the stirring mechanisms for 
cluster turbulence can be found in 
Refs.~\onlinecite{Subramanian_Shukurov_Haugen},~\onlinecite{Ensslin_Vogt_cores},~\onlinecite{Chandran_agn}. 

\begin{table}[t]
\caption{\label{tab_params} Cluster Parameters}
\begin{ruledtabular}
\begin{tabular}{llll}
Parameter & Expression       & Cool cores\footnotemark[1]      & Hot ICM\\
$T$       & observed         & $3\times10^7$~K                 & $10^8$~K\\
$n$       & observed         & $6\times10^{-2}~\text{cm}^{-3}$ & $10^{-3}~\text{cm}^{-3}$\\
$\vth$    & $(2T/m_i)^{1/2}$ & $700$~km/s                      & $1300$~km/s\\ 
$\nui$    & $1.5\,nT^{-3/2}$\footnotemark[2]       
                             & $5\times10^{-13}~\text{s}^{-1}$ & $2\times10^{-15}~\text{s}^{-1}$\\
$\mfp$    & $\vth/\nui$      & $0.05$~kpc                      & $30$~kpc\\
$\nupar$  & $\vth\mfp$       & $10^{28}~\text{cm}^2$/s         & $10^{31}~\text{cm}^2$/s\\
$\eta$    & $3\times10^{13}T^{-3/2}$\footnotemark[2]         
                             & $200~\text{cm}^2$/s             & $30~\text{cm}^2$/s\\
$U$       & inferred         & $250$~km/s                      & $300$~km/s\\
$L$       & inferred         & $10$~kpc                        & $200$~kpc\\
$L/U$     & inferred         & $4\times10^7$~yr                & $7\times10^8$~yr\\
$\Re$     & $UL/\nupar$      & $70$                            & $2$\\
$\Rm$     & $UL/\eta$        & $4\times10^{27}$                & $6\times10^{29}$\\
$\tvisc$  & $(L/U)\Re^{-1/2}$& $5\times10^6$~yr                & $5\times10^8$~yr\\
$\lvisc$  & $L\Re^{-3/4}$    & $0.4$~kpc                       & $100$~kpc\\ 
$\lres$   & $L\Rm^{-1/2}$    & $5000$~km                       & $8000$~km\\ 
$\Omeq$   & $e\Beq/cm_i$     & $0.3~\text{s}^{-1}$             & $0.04~\text{s}^{-1}$\\ 
$\rhoeq$  & $\vth/\Omeq$     & $3000$~km                       & $30,000$~km\\
$B_0$     & $\Beq\rhoeq/\mfp$& $5\times10^{-17}$~G             & $2\times10^{-19}$~G\\
$B_1$     & \eq{B1_def}\footnotemark[3]           
                             & $3\times10^{-14}$~G             & $2\times10^{-17}$~G\\
$B_2$     & \eq{B2_def}\footnotemark[3]           
                             & $8\times10^{-7}$~G              & $2\times10^{-7}$~G\\
$\Bvisc$  & $\Beq\Re^{-1/4}$ & $9\times10^{-6}$~G              & $4\times10^{-6}$~G\\
$\Beq$    & $(8\pi m_inU^2/2)^{1/2}$ 
                             & $3\times10^{-5}$~G              & $4\times10^{-6}$~G\\
$\betaeq$ & $8\pi nT/\Beq^2$ & $8$                             & $20$\\
$\lperp$  & $(B_2/\Beq)L$    & $0.2$~kpc                       & $7$~kpc\\
$\lB$     & observed         & $1$~kpc                         & $10$~kpc\\
\end{tabular}
\end{ruledtabular}
\footnotetext[1]{These numbers are based on the parameters for the Hydra A cluster given in 
Ref.~\onlinecite{Ensslin_Vogt_cores}.}
\footnotetext[2]{In these expressions, $n$ is in cm$^{-3}$, $T$ in Kelvin.}
\footnotetext[3]{We used $\alpha=3/2$ to get specific numbers, but the outcome is not 
very sensitive to the value of $\alpha$.}
\end{table}

While estimates of the turbulence parameters appear robust roughly 
to within an order of magnitude, a more quantitative set of numbers 
is elusive, partly because of the indirect and difficult nature 
of the observations, partly because the conditions vary both in 
different clusters and within each individual cluster. Here we shall adopt two 
fiducial sets of parameters: one for cool cores and one for 
the bulk of hot cluster plasma. These are given in \tabref{tab_params} 
(along with some theoretical 
quantities that will arise in \secref{sec_dynamo} and \secref{sec_sat}). 
They will allow us to make  
estimates that will have the virtue of being consistent and 
systematic but must not be interpreted as precise quantitative 
predictions. They are representative of the range of conditions 
that can be present in clusters. 
The turbulence is assumed to be stirred at 
the outer scale $L$ (with rms velocity $U$ at this scale) 
and to have a Kolmogrov-type cascade below this scale. 
The small-scale cutoff is determined by 
the microphysical properties of the cluster plasma. 
In \tabref{tab_params}, we give the value 
of the particle mean free path $\mfp$, which can be used in 
a naive estimate of viscosity: $\nuicm\sim\vth\mfp$, where $\vth=(2T/m_i)^{1/2}$ 
is the ion thermal speed. We see that this 
gives fairly low values for the Reynolds number, $\Re\sim UL/\nuicm$. 
It is this feature of the ICM that continues to fuel doubts 
about its ability to support turbulence, at least in the strict, 
hydrodynamic high-Reynolds-number sense.\footnote{These doubts 
are reinforced by the similarity between the inferred flow 
patterns associated with rising bubbles and known 
such patterns in viscous, rather than 
turbulent, fluid flows.\cite{Fabian_etal_Perseus2}} 
However, one should be cognizant of the fact 
that whatever type of turbulence might exist in the cluster plasma, 
it is certainly not hydrodynamic, because this plasma is highly 
electrically conducting\footnote{An estimate of resistivity $\eta$ based on the standard 
Spitzer formula leads to the enormous values of the magnetic 
Reynolds number, $\Rm\sim UL/\eta$, given in \tabref{tab_params}.} 
and magnetized. The presence of the magnetic fields 
not only has a dynamical effect on the turbulence (due to the 
action of the Lorentz force, the medium acquires a certain elastic 
quality), but also changes the transport properties of the plasma itself: 
the viscosity, in particular, becomes strongly anisotropic.\cite{Braginskii} 
These issues will constitute the main subject of this paper, 
but first let us briefly describe what is known about magnetic fields 
in clusters. 

\section{Magnetic fields}
\label{sec_mag}

The first observed signature of cluster magnetic fields was the 
diffuse synchrotron radio emission in the Coma cluster detected 
in 1970.\cite{Willson} Starting from early 1990's, increasingly 
detailed measurements of the Faraday Rotation in the 
emission from intracluster radio sources have made possible 
quantitative estimates of the magnetic field strength and scales 
in a large number of clusters.\cite{Kronberg,Carilli_Taylor,Govoni_Feretti} 
Randomly tangled magnetic fields with rms strength 
of order $\Brms\sim1-10~\mu$G are consistently found,  
with the fields in the cool cores of the cooling-flow clusters
somewhat stronger than elsewhere. This is fairly close to the 
value $\Beq$ that corresponds 
to magnetic energy equal to the energy of turbulent motions 
(see \tabref{tab_params}). 
Thus, the magnetic field must be dynamically important. 
The estimates for the tangling scale $\lB$ of the field 
are usually arrived at by assuming 
that direction reversals along the line of sight 
(probed by the Faraday Rotation measure) can be described 
as a random walk with a single step size equal to $\lB$ 
(the estimate of $\Brms$ is obtained in conjunction with 
this model). This gives $\lB\sim1-10$~kpc. 

The single-scale model is almost certainly not a correct description 
on any but a very rough level. Fortunately, much more detailed 
information on the spatial structure of the cluster fields is 
accessible. First, using certain statistical assumptions 
(most importantly, isotropy), it is possible 
to compute magnetic-energy spectra from the maps of the Faraday Rotation 
measure associated with extended radio sources (the radio lobes of the jets 
emerging from the AGN --- these can be as large as $\sim10^2$~kpc 
across).\cite{Ensslin_Vogt_method,Vogt_Ensslin1} 
This has been done most thoroughly for a radio lobe located 
in the cool core of the Hydra A cluster.\cite{Vogt_Ensslin2} 
The spectrum has a peak at $k\simeq2~\text{kpc}^{-1}$ followed 
by what appears to be a power tail consistent with $k^{-5/3}$ 
down to the resolution limit of $k\simeq10~\text{kpc}^{-1}$. 
The rms magnetic field strength is $\Brms=7\pm2~\mu$G. 

The second source of information on the cluster field structure 
is the polarized synchrotron emission, which probes the magnetic field 
in the plane perpendicular to the line of sight.\cite{Burn} Such data, 
while widely used for Galactic magnetic field studies,\cite{Haverkorn_Katgert_deBruyn} 
has until recently not been available for clusters. This is now 
changing: the first analysis of polarized emission from a radio relic in 
the cluster A2256 reveals the presence of magnetic filaments with field reversals 
probably on $\sim20$~kpc scale, which, however, is dangerously 
close to the resolution scale.\cite{Clarke_Ensslin} This data is 
representative of the situation in the bulk of the ICM, rather than 
in the cores. Statistical analysis of such data will make possible 
quantitative diagnosis of the field structure and its dynamical role.\cite{EWVS_bologna} 

Thus, our knowledge of the magnetic fields in clusters, while far 
from perfect, is more direct and more detailed than that of 
the turbulent motions of the ICM. It is also due to improve dramatically 
with the arrival of new radio telescopes such as LOFAR and SKA.\footnote{http://www.lofar.org/; http://www.skatelescope.org/.}

\section{Plasma physics} 
\label{sec_plasma}

The key property of the ICM as plasma is that it is only weakly collisional 
and magnetized: given the observed values of the magnetic field,
the ion gyroradius is $\rho_i\sim10^4$~km, which is much smaller 
than the mean free path. As $\rho_i\ll\mfp$ already for 
dynamically very weak fields ($B\gg B_0$, see \tabref{tab_params}), 
this is true both in the observed present state of the ICM 
and during most of its hypothetical past, when the magnetic field was 
being amplified from some weak seed value.
In a plasma with $\rho_i\ll\mfp$, the equations 
for the flow velocity $\vu$ and for the magnetic field $\vB$ 
may be written in the following form, valid 
at time scales $\gg\Omega_i^{-1}$ 
($\Omega_i=eB/m_ic$ is the ion cyclotron frequency) 
and spatial scales $\gg\rho_i=\vth/\Omega_i$, 
\bea
\label{u_eq_brag}
\rho\,{d\vu\over dt}\!\! &=&\!\! -\vdel\!\lt(\pperp+{B^2\over2}\rt)
+ \vdel\!\!\cdot\!\lt[\vb\vb\lt(\pperp-\ppar + B^2\rt)\rt],\ \\
{d\vB\over dt}\!\! &=& \!\!\vB\cdot\!\vdel\vu - \vB\vdel\cdot\vu,
\label{ind_eq}
\eea
where $d/dt=\dd/\dd t + \vu\cdot\vdel$ is the convective derivative, 
$\pperp$ and $\ppar$ are plasma pressures perpendicular 
and parallel to the local direction of the magnetic field $\vB$, 
respectively, $\vb=\vB/B$, the factor of $1/\sqrt{4\pi}$ has been absorbed 
into $\vB$, and the resistive term has been omitted 
in \eq{ind_eq} in view of the tiny value of the resistivity. 
The turbulent motions in clusters are subsonic ($U<\vth$), 
so we may take $\vdel\cdot\vu=0$ and set $\rho=1$. The magnetic field 
is, thus, in units of velocity, pressure in units of velocity squared.  

%If we are interested in subsonic motions, $\vdel(\pperp+B^2/2)$ in 
%\eq{u_eq_brag} can be found from the incompressibility condition 
%$\vdel\cdot\vu=0$ and the only quantity still to be determined 
%is $\pperp-\ppar$. 

The proper way to compute $\pperp$ and $\ppar$ is by 
a kinetic calculation. In the collisional limit, this 
was done in Braginskii's classic paper.\cite{Braginskii} 
It is instructive to obtain his result 
in the following heuristic way that highlights the 
physics behind the formalism.\cite{SCKHS_brag} 
Charged particles moving in a magnetic field conserve  
their first adiabatic invariant $\mu=m_i\vperp^2/2B$.
When $\mfp\gg\rho_i$, this conservation is only weakly broken by collisions. 
As long as $\mu$ is conserved, any change 
in the field strength causes a proportional change in $\pperp$: 
summing up the first adiabatic invariants of all 
particles, we get $\pperp/B=\const$. Then 
\bea
{1\over\pperp}{d\pperp\over dt} 
\sim {1\over B}{dB\over dt} 
-\nui\,{\pperp-\ppar\over\pperp},
\label{dlnB_approx}
\eea
where the second term on the right-hand sight represents 
the relaxation of the pressure anisotropy 
at the ion collision rate $\nui\sim\vth/\mfp$.\footnote{This is 
only valid if the characteristic parallel scales $\kpar^{-1}$ 
of all fields are larger than $\mfp$. In the collisionless regime, 
$\kpar\mfp\gg1$, we may assume that the 
pressure anisotropy is relaxed in the time particles streaming 
along the field cover the distance $\kpar^{-1}$: this entails 
replacing $\nui$ in \eq{dlnB_approx} by $\kpar\vth$.} 
Using \eq{ind_eq} with $\vdel\cdot\vu=0$ 
and balancing the terms in the rhs of \eq{dlnB_approx}, we get 
\bea
\pperp-\ppar \sim \nupar\,{1\over B}{dB\over dt} = \nupar\vb\vb:\vdel\vu, 
\label{Brag_visc}
\eea
where $\nupar\sim \pperp/\nui\sim\vth\mfp$ is the ``parallel viscosity.'' 
\Eq{Brag_visc} turns out to be quantitatively correct provided the value of $\nupar$ 
with the correct prefactor calculated by Braginskii\cite{Braginskii} is used. 
Thus, the emergence of the pressure anisotropy is a natural 
consequence of the changes in the magnetic-field strength and vice 
versa.\footnote{The anisotropy is small: 
$\Delta=(\pperp-\ppar)/\pperp\sim (U/\vth)^2\Re^{-1/2}$. 
It turns out that $(U/\vth)\Re^{-1/4}$ is the natural small parameter 
that can be used to develop a reduced kinetic theory 
for the cluster plasma.\cite{SCKHS_brag}} 

The energy conservation law based on \eqsdash{u_eq_brag}{ind_eq} 
with $\vdel\cdot\vu=0$ and on \eq{Brag_visc} is
\bea
\nonumber
{d\over dt}\lt({\usq\over2} + {\Bsq\over2}\rt) 
&=& - \nupar\bl<|\vb\vb:\vdel\vu|^2\br>\\ 
&=& - \nupar\lt<\lt({1\over B}{dB\over dt}\rt)^2\rt>.
\label{energy_cons}
\eea
Thus, the Braginskii viscosity only dissipates 
such motions that change the strength of the 
magnetic field. Motions that do not affect $B$ are allowed 
at subviscous scales. 
In the weak-field regime, these motions take the form of plasma 
instabilities. When the magnetic field is strong, a cascade of 
Alfv\'en waves can be set up below the viscous scale. 
Let us elaborate. 

The simplest way to see that pressure anisotropies 
lead to instabilities is as follows.\cite{SCKHS_brag} Imagine that 
the large-scale energy sources stir up a ``fluid'' 
turbulence with $\vu$, $\pperp$, $\ppar$, $\vB$ at time and 
spatial scales above viscous. 
Would such a solution be stable with respect to much higher-frequency and  
smaller-scale perturbations? Linearizing \eq{u_eq_brag} 
and denoting perturbations by $\delta$, we~get 
\bea
\nonumber
\!\!\!\!-i\omega\dvu\!\! &=&\!\! -i\vk\lt(\delta\pperp + B\delta B\rt) 
+ \lt(\pperp-\ppar + B^2\rt)\delta\vK\\ 
&&\!\!+\,\, i\vb\,\kpar\!\lt[\delta\pperp-\delta\ppar - \lt(\pperp - \ppar - B^2\rt) 
{\delta B\over B}\rt],\ 
\label{du_eq}
\eea
where $\vK=\vb\cdot\vdel\vb$ is the curvature of the field. Using \eq{ind_eq},  
$\delta\vK=\kpar^2\dvuperp/i\omega$. %and $\delta B/B\sim -\kpar\dupar/\omega$. 
We can immediately separate the Alfv\'en-wave-polarized perturbations 
($\dvu\propto\vk\times\vb$), for which we obtain the dispersion relation 
\bea
\omega = \pm\kpar\lt(\pperp-\ppar+B^2\rt)^{1/2}. 
\label{firehose}
\eea
When $\ppar-\pperp>B^2$, $\omega$ is purely imaginary and we have 
what is known as the firehose instability.\cite{Rosenbluth,Chandrasekhar_Kaufman_Watson,Parker_firehose,Vedenov_Sagdeev} 
The growth rate of the instability is $\propto\kpar$, which means 
that the fastest-growing perturbations are at scales far below 
the viscous scale or the mean free path. Therefore, 
\eqsdash{u_eq_brag}{ind_eq} are ill posed wherever $\ppar-\pperp>B^2$. 
To take into account the instability and its impact on the large-scale 
dynamics, a kinetic, rather than fluid, description must be adopted. 
A linear calculation based on the hot plasma dispersion relation shows that 
the instability growth rate peaks at the ion gyroscale, $\kpar\rho_i\sim1$. 
While the firehose instability occurs in regions where 
the magnetic-field strength is decreasing [\eq{Brag_visc}], 
a kinetic calculation of $\delta\pperp$ and $\delta\ppar$ in \eq{du_eq} 
yields another instability, called the mirror mode,\cite{Parker_firehose} 
that is triggered wherever the field is increasing ($\pperp>\ppar$).
Its growth rate also peaks at $\kpar\rho_i\sim1$. 

To sum up briefly, external energy sources 
drive random motions, which change the field [\eq{ind_eq}], 
which gives rise to pressure anisotropies 
[\eq{Brag_visc}], which trigger the instabilities. 
The latter are stabilized when $B^2>|\pperp-\ppar|$ and the 
firehose fluctuations, in particular, mutate into Alfv\'en waves, 
which can cascade without being affected by collisions 
all the way to the ion gyroscale.\cite{SCDHHQ_gk} 
We shall revisit the subject of the Alfv\'en-wave cascade 
in \secref{sec_sat}, but first let us see what the extraordinarily 
unstable nature of the ICM in the weak-field regime implies. 

What it implies is, in fact, not altogether clear, as the quantitative 
theory of the evolution of the instabilities beyond the linear stage 
is a difficult task.\footnote{In space physics, where the 
importance of the plasma instabilities driven by pressure anisotropies 
of what in their case is an almost entirely collisionless plasma 
has long been understood.\cite{Gary_book} A vast but inconclusive 
literature exists on this subject.} 
While the standard quasilinear scheme can be 
implemented more or less rigorously in the limit of small anisotropy 
for particular cases in which the 
instability growth rates peak at scales much larger than the ion gyroscale 
(specifically, when $\kperp=0$ for the firehose and $\kperp\gg\kpar$ 
for the mirror),\cite{Shapiro_Shevchenko} it is not clear that this 
is relevant, because in the general case, the growth rates peak at 
$k\rho_i\sim1$. The difference between this situation and the case 
of $k\rho_i\ll1$ is qualitative: for the fluctuations at the 
gyroscale, the $\mu$ conservation is broken, so they can give rise 
to effective particle scattering, while the quasilinear saturation 
of the fluctuations with $\kpar\rho_i\ll1$ happens essentially because 
$\delta\pperp$ and $\delta\ppar$ offset the initial pressure 
anisotropy.\cite{Quest_Shapiro} 

\begin{figure}[t]
\centerline{\psfig{file=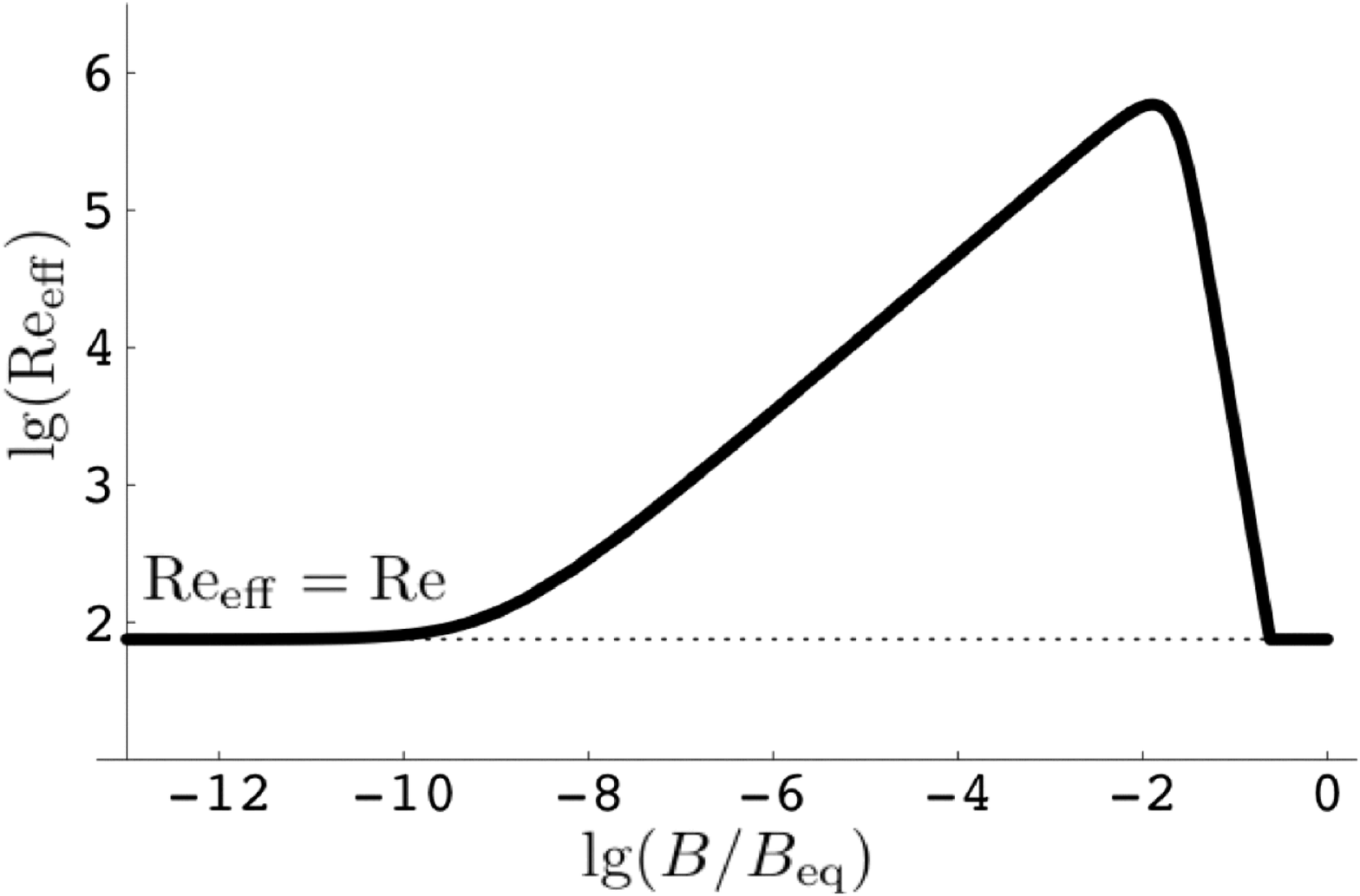,width=8cm}} 
%\vskip-0.5cm
\caption{\label{fig_Reeff} $\Reeff(B)$: solution of \eq{Reeff_eq} calculated 
numerically for the cool-core parameters of \tabref{tab_params}.}
%\vskip-0.25cm
\end{figure}

Here we would like to leapfrog the 
rather forbidding task of constructing a quasilinear theory based on 
the hot plasma dispersion relation with $k\rho_i\sim1$ and instead 
simply assume that the istabilities saturate with 
${\la\delta B^2\ra/B^2} \sim \bl(|\Delta|-{2/\beta}\br)^{\alpha_1}$,
where $\Delta=(\pperp-\ppar)/\pperp$, $\beta=2\pperp/B^2$, 
and $\alpha_1$ is some positive power. 
We further assume that the effect of the saturated instabilities 
is to scatter particles at the rate $\nusc \sim \gmax \la\delta B^2\ra/B^2$, 
where $\gmax \sim \bl(|\Delta|-{2/\beta}\br)^{\alpha_2}\Omega_i$ 
is the maximum growth rate of the instabilities ($\alpha_2=1$ for the mirror 
and $\alpha_2=1/2$ for the firehose [\eq{firehose}]). 
We may then construct the effective collision rate  
\bea
\label{nueff_def}
\nueff \sim \nui + \nusc 
\sim \nui + \lt(|\Delta|-{2\over\beta}\rt)^{\alpha}\Omega_i,
\eea 
where $\alpha=\alpha_1+\alpha_2>0$. This changes 
the characteristics of the turbulence: the effective  
mean free path of the particles is $\mfpeff\sim \vth/\nueff$, the 
effective (parallel) viscosity of the ICM is $\nupareff\sim \vth\mfpeff$ and, therefore, 
the effective Reynolds number is 
\bea
\label{Reeff_def}
\Reeff\sim {UL\over\nupareff}\sim {UL\over\vth^2}\,\nueff.
\eea 
On the other hand, using \eq{Brag_visc} with effective viscosity 
$\nupareff$ and $|\vdel\vu|\sim (U/L)\Reeff^{1/2}$, we get 
\bea
\label{Delta_def}
|\Delta|\sim \lt({U\over\vth}\rt)^2\Reeff^{-1/2}. 
\eea
From \eqsdash{nueff_def}{Delta_def}, we assemble a (model) equation 
for the effective Reynolds number:
\bea
%\!\!\!\!\Reeff = \Re + {L\over\rhoeq}\lt[\lt({U\over\vth}\rt)^2\!\!\! 
%{1\over\Reeff^{1/2}} - {2\over\beta}\rt]^\alpha\!\!\!{1\over\sqrt{\beta}},
\!\!\Reeff = \Re + {L\over\rhoeq}\lt({U\over\vth}\rt)^{\!\!2\alpha+1}\!\!\! 
\lt({1\over\sqrt{\Reeff}} - {2B^2\over\Beq^2}\rt)^\alpha\!\!\!{B\over\Beq},\ 
\label{Reeff_eq}
\eea
where $\Re$ is the original Reynolds number based on collisions 
and $\rhoeq$ is the ion gyroradius for $B=\Beq$. 
The assumption that $\Reeff$ adjusts to the value of $B$ 
instantaneously is justified by the fact that the instability growth 
rates are much faster than all other relevant time scales. 

\Eq{Reeff_eq} models qualitatively the assumed 
outcome of an as yet inexistent proper theory of the viscosity of magnetized 
ICM. Its solution is plotted in \figref{fig_Reeff}. 
We shall use this model shortly in our discussion of the 
fluctuation dynamo in clusters. 

\section{Fluctuation dynamo} 
\label{sec_dynamo}

\begin{figure}[t]
\centerline{\psfig{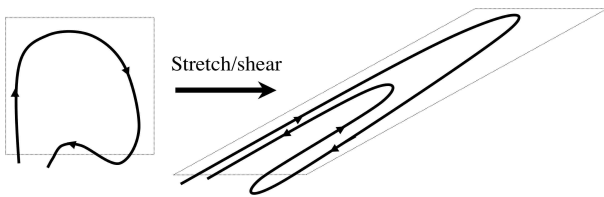}} 
%\vskip-0.5cm
\caption{\label{fig_stretch} The mechanism of the fluctuation dynamo.}
%\vskip-0.25cm
\end{figure}

In \secref{sec_mag}, we reviewed the observational evidence that testified 
to the presence of a dynamically significant randomly tangled magnetic field 
in clusters. What is the origin of this field? 
There are numerous physical reasons to expect that a certain amount 
of seed magnetic energy predates structure formation and 
was, therefore, already present at the birth of 
clusters.\cite{Grasso_Rubinstein,Gnedin_Ferrara_Zweibel} 
Typical values given for the strength of such field are 
in the range of $\Bseed\sim10^{-21}-10^{-17}$, although 
this may be an underestimate.\cite{Banerjee_Jedamzik}  
It then falls to the random motions of the cluster plasma 
to amplify the field to its observed magnitude of a 
few $\mu$G.\footnote{Several authors have argued that no amplification 
is, in fact, necessary either because the seed 
fields may have already been strong enough to account, after 
compression of the cosmological plasma into cluster, for the observed 
magnitude of the cluster field,\cite{Banerjee_Jedamzik} 
or because the field could be generated in AGNs and then ejected 
into the ICM.\cite{Kronberg_etal_agns} We forego the discussion 
of these possibilities, primarily because we find the idea that 
the ICM turbulence will produce the right amount of magnetic energy 
independently of its history or external circumstances more 
appealing on a fundamental physics level and, indeed, intuitively 
supported by the fact that the energy density of the observed field 
is close to the energy density of the fluid motions.} 
This, indeed, they should be able to do 
by means of the fluctuation (or small-scale) dynamo mechanism: 
the random stretching of the field. It is a fundamental 
property of a succession of random (in time) linear shears that 
it leads on the average to exponential growth of the energy of 
the magnetic field frozen into the 
medium.\cite{Batchelor_dynamo,Zeldovich_etal_linear,Almighty_Chance} 
The rate of growth is roughly equal to the rate of strain 
(shear, or stretching rate) of the random flow. 
While the mathematical theory of this process can be nontrivial,\cite{Arnold_Khesin} 
the physics of it is basically illustrated by \figref{fig_stretch}. 

In Kolmogorov turbulence, the rate of strain is dominated by the 
viscous scale, so $|\vdel\vu|\sim \tvisc^{-1}\sim(U/L)\Re^{1/2}$. In fact, what is 
relevant for the growth of the magnetic field is not the full rate-of-strain 
tensor but its ``parallel'' component, $\vb\vb:\vdel\vu$. Since this is exactly 
the type of motion damped by Braginskii viscosity [see \eqsdash{Brag_visc}{energy_cons}], 
we can, for the purposes of the fluctuation dynamo, 
ignore any subviscous-scale velocity fluctuations. 
Thus, the magnetic field should grow according to\footnote{Note that 
turbulence in the sense of a broad inertial range is not 
required for the fluctuation dynamo. The viscous-scale motions, 
which dominantly stretch the field, are random but spatially smooth. 
Whether the viscous scale $\lvisc\sim L\Re^{-3/4}$ is much smaller than 
the outer scale $L$ is inessetial as long as the motions are 
random --- not a problem in clusters, given the vigorous 
random stirring. Thus, the Reynolds number in all of our 
calculations need not be large. Numerical simulations of a randomly 
stirred magnetohydrodynamic (MHD) fluid with $\Re$ in the range $1-10^3$ 
confirm the insensitivity of the dynamo effect to the value of $\Re$
\cite{SCTMM_stokes,Haugen_Brandenburg_Dobler}.} 
\bea
\label{B_growth}
{1\over B}{dB\over dt} = \vb\vb:\vdel\vu \sim {U\over L}\,\Re^{1/2}. 
\eea
In order to be useful, this amplification has to be done 
over a time that is shorter than the typical age of the clusters, 
i.e., a few Gyr. As $L/U\sim1$~Gyr and as 
the Reynolds number based on collisional ICM viscosity is quite low, 
it has been a concern of many authors that the cluster life time might not 
be (or, possibly, is only just) sufficient to bring the field to 
its current strength\cite{Jaffe,Roland,Ruzmaikin_Sokoloff_Shukurov,DeYoung,Goldshmidt_Rephaeli,SanchezSalcedo_Brandenburg_Shukurov,Subramanian_Shukurov_Haugen} 
(this obviously depends on the magnitude of the seed field). 
Similar issues arise in simulations of the growth of magnetic fields in clusters.\cite{Roettinger_Stone_Burns,Dolag_Bartelmann_Lesch1,Dolag_Bartelmann_Lesch2,Dolag_etal_mag,Brueggen_etal_mag} 
However, if \eq{Reeff_eq} is a good model of the ICM viscosity, 
the value of the Reynolds number based on the collisional 
viscosity has to be amended when the ICM is magnetized:  
since the prefactor $L/\rhoeq$ in the second, magnetic-field dependent 
term in \eq{Reeff_eq} is extremely large (see \tabref{tab_params}), 
this term will dominate already for 
very small values of the magnetic field strength, namely, for $B>B_1$, where 
\bea
\label{B1_def}
B_1 = \Beq\,{\rhoeq\over L}\biggl({\vth\over U}\biggr)^{1+2\alpha}\Re^{1+\alpha/2}.
\eea
$B_1$ is typically so small (\tabref{tab_params}) that 
the seed field in clusters may already exceed this value. 
Thus, the effective Reynolds number of the ICM will be much larger than 
the one based on collisions. 
Substituting $\Reeff$ instead of $\Re$ into \eq{B_growth}, 
we see that, since $\Reeff\propto B^{2/(2+\alpha)}$ in the weak-field regime,  
the dynamo is self-accelerating and the 
field will increase not exponentially but explosively: 
\bea
B(t) \sim {B(0)\over \lt(1-t/t_c\rt)^{2+\alpha}},
\eea
where $B(0)\sim$ the greater of $\Bseed$ and $B_1$ and 
$t_c=(2+\alpha)(L/U)\Re^{-1/2}\lt[B_1/B(0)\rt]^{1/(2+\alpha)}$ 
is at most (for $\Bseed<B_1$) the viscous turnover time $\tvisc$ 
associated with the collisional $\Re$. The explosive stage continues until 
the amplified field starts suppressing the instabilities, i.e., 
when $\beta$ drops to values comparable to $(\vth/U)^2\Reeff^{1/2}$. 
This happens at $B\sim B_2$, where 
\bea
\label{B2_def}
B_2 = \Beq\lt[\biggl({\vth\over U}\biggr)^{1+2\alpha}
{\rhoeq\over L}\rt]^{1/(5+2\alpha)}.
\eea
Thus, we have a mechanism that amplifies the field by many orders of 
magnitude from any strength above $B_1$ to $B_2$ in finite, 
cosmologically short, time $\sim t_c$. 
The value of $B_2$ turns out to be only just over an order of magnitude 
below $\Beq$ (see \tabref{tab_params}). 

Further growth of the field is algebraically slow 
($B\sim t^{1/2}$), but it does not have to go on for a very long time 
because $B_2$ is already quite close to the observed field strength. 
To be precise, there are two algebraic regimes. 
During the first, $\Reeff$ is still controlled by the second term in 
\eq{Reeff_eq} as $B$ hovers just below $\Beq\Reeff^{-1/4}$ while 
$B$ is increasing and $\Reeff$ is decreasing. Eventually, 
$B\sim\Bvisc=\Beq\Re^{-1/4}$ and $\Reeff$ is returned back 
to $\Re$ (plasma instabilities are suppressed).\footnote{This, 
effectively, is the regime assumed by Sharma {\em et al.}\cite{Sharma_etal} 
in their numerical simulations of the magnetorotational instability 
in a collisionless plasma. Their set of 
fluidlike equations incorporates anisotropic pressure and an effective 
collision rate that enforces the marginal state of the plasma instabilities: 
$(\vth/U)^2\Reeff^{1/2}\simeq\beta$ in our notation.} 
As this is also the field strength at which the field has energy comparable to the 
energy of the viscous-scale motions, any further growth of the 
field is a nonlinear process, in which the back reaction of the 
field on the flow has to be taken into account. This can be done 
by assuming that, as the field grows, it can no longer be 
stretched by motions whose energy it exceeds and that, therefore, 
at any given time, the dominant stretching is exerted by motions 
whose energy is equal to that of the field. Denoting their 
velocity and scale by $u_l$ and $l$ and using 
$u_l^2\sim B^2$, we have\cite{SCHMM_ssim,SCTMM_stokes}  
\bea
{d\over dt} B^2 \sim {u_l\over l}\,B^2 \sim {u_l^3\over l} \sim \epsilon 
= \const,
\label{growth_ssim}
\eea
where $\epsilon$ is the energy flux through scale $l$, which, 
by Kolmogorov's assumption, is independent of $l$. Thus, 
$B\sim (\epsilon t)^{1/2}$, the second algebraic regime. 
Obviously, this can only go on until $l\sim L$, $B\sim\Beq$, 
and no further amplification is possible. 

\begin{figure}[t]
\centerline{\psfig{file=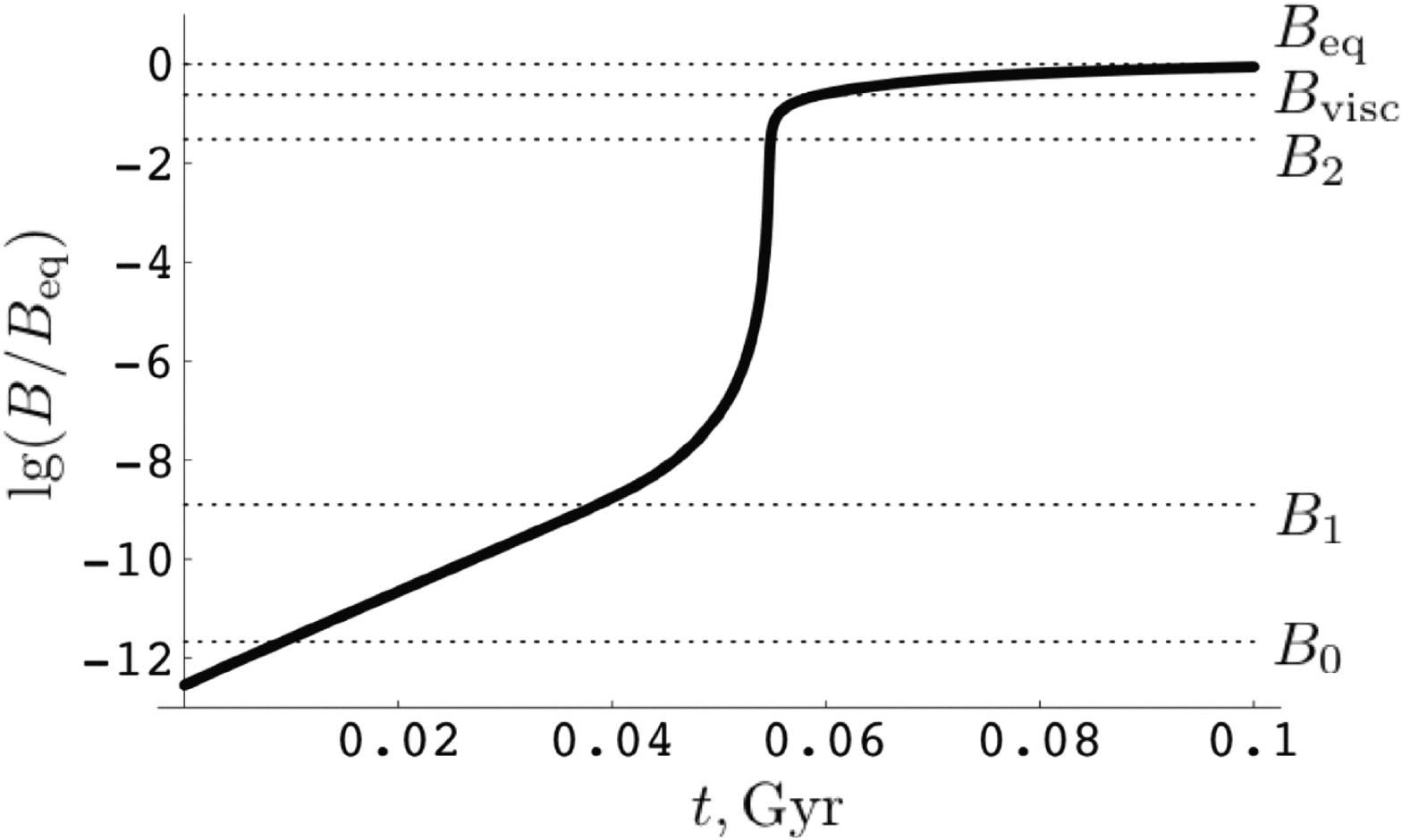,width=8cm}} 
%\vskip-0.5cm
\caption{\label{fig_Bvst} Evolution of the magnetic-field strength for the 
cool-core parameters of \tabref{tab_params}.}
%\vskip-0.25cm
\end{figure}

The time evolution of the field is illustrated 
in \figref{fig_Bvst}, where we plot 
the result of a numerical integration of \eq{B_growth} 
with $\Re$ replaced by $\Reeff$. To implement the idea 
of the nonlinear suppression of stretching by motions whose 
energy is smaller than $B^2$, we have amended the definition 
of $\Reeff$ in the following simple way\cite{SCHMM_ssim} 
\bea
\Reeff^{\rm (nonlin)} = {\Reeff\over 1-A} 
\lt[{1\over(1+\Reeff^{1/2}B^2/\Beq^2)^2}-A\rt],
\eea
where $A=1/\bl(1+\Reeff^{1/2}\br)^2$, so that 
$\Reeff^{\rm (nonlin)}=\Reeff$ when $B\to0$ 
and $\Reeff^{\rm (nonlin)}\to0$ when $B\to\Beq$. 
$\Reeff$ is determined for each value of $B$ by \eq{Reeff_eq}. 

\section{Magnetized plasma turbulence} 
\label{sec_sat}

Thus, the clusters should have no trouble developing magnetic fields 
of observed strength over just a fraction of their life 
time.\footnote{This means, in particular, that young clusters 
should already have dynamically significant magnetic fields.} 
This, however, is only one part of the problem. The other, much 
more difficult, part is to determine the spatial structure of the field. 
Theory\cite{Ott_review,SCMM_folding} and numerical 
simulations\cite{SCTMM_stokes,Brandenburg_Subramanian} 
of the fluctuation dynamo in a conventional MHD formulation with 
fixed isotropic viscosity and much smaller resistivity 
($\Rm\gg\Re$) produce magnetic fields arranged 
in folded flux sheets whose parallel (locally to themselves) scale 
is similar to the scale of the velocity that does the stretching, while 
the field reversal scale in the direction perpendicular to itself 
is the resistive scale. \Figref{fig_stretch} illustrates 
why such a folded structure must be a generic outcome of 
random stretching. Less obviously, it turns out\cite{SCTMM_stokes} that 
the folded structure is a feature not only of the kinematic growth stage 
of the dynamo but also of its saturated state (\figref{fig_slice}). 
In the latter case, the fold length is, thus, $\lpar \sim L$, while 
the reversal scale is $\lperp\sim\lres\sim L\Rm^{-1/2}$. The magnetic-energy spectra 
of such fields appear to peak around $k\sim 1/\lperp$. 

\begin{figure}[t]
\centerline{\psfig{file=u_z480_square.epsf,width=4cm}
\hskip0.25cm
\psfig{file=B_z480_square.epsf,width=4cm}} 
%\vskip-0.5cm
\caption{\label{fig_slice} Cross sections of $|\vu|$ (left) 
and $|\vB|$ (right) in the saturated state of a simulation 
with $\Re\simeq100$, $\Rm=1000$ 
(run B of Ref.~\onlinecite{SCTMM_stokes}).}
%\vskip-0.25cm
\end{figure}

This prediction is 
clearly not in agreement with what seems to be the observationally 
supported picture of the magnetic fields in clusters. While the typical 
scale $\lB$ of the cluster fields is significantly smaller than 
the outer scale $L$ of the turbulent velocities, it is certainly not 
the resistive scale --- at least not the one based on the standard 
Spitzer resistivity. What then determines the reversal 
scale in clusters? While the definite solution of this puzzle remains 
elusive, we offer the following qualitative argument that 
at least constrains the answer. 

%An essential feature of the folded field structure produced by 
%random stretching is that field-line curvature $K=|\vb\cdot\vdel\vb|$
%and the field strength are anticorrelated so that roughly 
%$B^2K\sim\const$ (this means, incidentally,  
%that the tension force exerted 
%by the field is approximately the same everywhere). 

Examining again \figref{fig_stretch}, it should be obvious that if the 
folded magnetic field is to reverse its direction, it must turn a corner 
somewhere and that in that corner, the field should be weaker than 
in the straight segments of the fold. Furthermore, as there are 
antiparallel fields in the straight segments themselves, there must be 
layers of weak field between them (assuming that folds are 
flux sheets: i.e., that the antiparallel fields in \figref{fig_stretch} 
are, indeed, approximately aligned). These are, in fact, also of the ``corner'' 
type: this becomes clear if the fold depicted in \figref{fig_stretch} is thought 
of as a fragment of a larger nested folded structure formed by 
repeated stretching/shearing and bending. 
This view of the dynamo-generated fields as either strong and straight 
or weak and curved is confirmed in numerical 
simulations, which find an almost perfect anticorrelation between the 
field's strength and its curvature.\cite{SCTMM_stokes} 
While the exact quantitative relation between the field strength and 
its scale may be a nontrivial issue, here we shall continue in the spirit 
of intuitive reasoning and argue that by flux conservation, 
$\Bstr/\Bcorner\sim\lpar/\lperp$ (this again relies on assuming that folds are 
flux sheets). Thus, the larger is the aspect ratio $\lpar/\lperp$, 
the larger is the contrast between the strong field 
in the straight segments of the folds and the weak field in the corners. 
We assume that the rms value of the field is determined by the 
straight segments because these are the fields that are stretched 
by turbulence. It is their growth that we studied in \secref{sec_dynamo}. 
In the saturated state, we expect $\Bstr\sim\Beq$. 
For such a strong field, the plasma instabilities are suppressed. 
It is intuitively clear that they must be suppressed in the regions 
of the weak field as well, i.e., the field there cannot be weaker than 
$B_2$ [\eq{B2_def}]. Indeed, as we saw in \secref{sec_dynamo}, 
the explosive dynamo mechanism brings any field up to this value 
nearly instantaneously (for $B>B_2$, the growth is much slower). 
We may conjecture that the maximum aspect ratio of the folds is set by the maximum 
contrast in the field strength $\lpar/\lperp\sim\Beq/B_2$.\footnote{This does 
not mean that cluster fields must be volume filling. Indeed, a real 
cluster is probably a patchwork of turbulent and quiescent regions 
(or rather regions with widely varying rms rates of strain) 
rather than a volume homogeneously filled with turbulence. 
Furthermore, the turbulent dynamo only produces magnetic fields 
everywhere on the average (in time), while any particular snapshot 
is very intermittent\cite{SCTMM_stokes} --- see, e.g., \figref{fig_slice}.} 
Substituting the numbers, we find (\tabref{tab_params}) that this prediction 
gives the field reversal scale no smaller than a few per cent of the outer 
scale $L$ (taking $\lpar\sim L$). This is in passable agreement with 
the observational evidence, which is the best one can expect, given the highly 
imprecise nature of our argument, of most observational inferences, and 
of the definitions of such quantities as $\lB$, $\lperp$, $\lpar$ and $L$. 
For comparison of our model with observational data for a number 
of individual clusters, see Ref.~\onlinecite{Ensslin_Vogt_cores}. 

\begin{figure}[t]
\centerline{\psfig{file=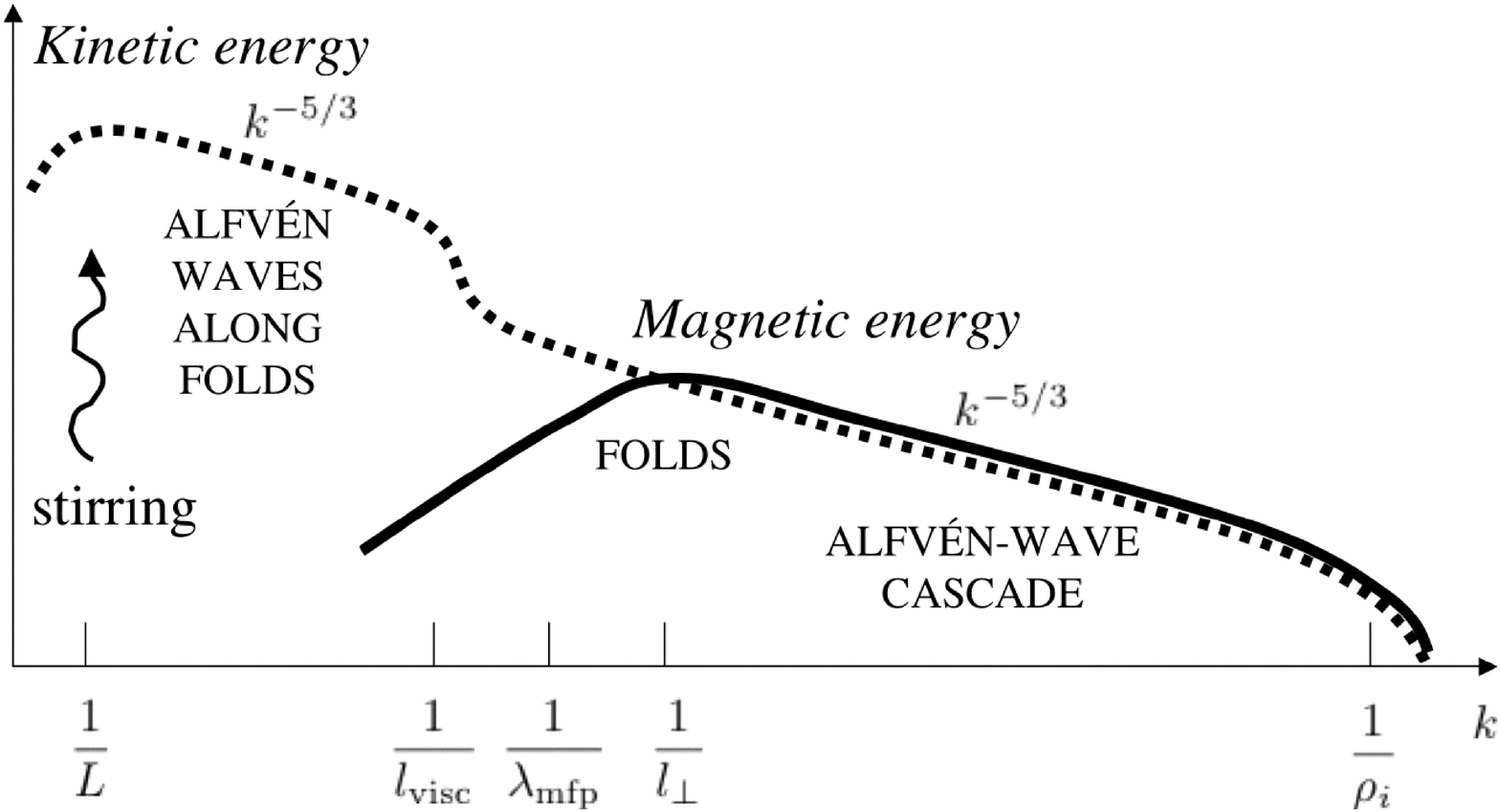,width=8cm}} 
%\vskip-0.5cm
\caption{\label{fig_turb} A schematic illustration of the structure of 
cluster turbulence proposed in \secref{sec_sat}.}
%\vskip-0.25cm
\end{figure}

It is fair to acknowledge that the above argument, while providing 
a useful constraint, falls short of a satisfactory explanation of 
the field structure. 
One might argue that if, in the course of the turbulent stretching/shearing of the 
field, a region of field strength below $B_2$ appears (as explained above, in the 
corner of a fold), $\Reeff$ there becomes very large and a 
localized spot of high-Reynolds-number turbulence is formed. 
This should have two principal effects. The first is akin to that 
of a locally enhanced turbulent resistivity, so the field that 
violates our constraint is continuously destroyed. 
The second is a burst of explosive fluctuation dynamo in the spot, 
which produces more folded field with $B>B_2$ and thus shuts itself down. 
These folds are then further stretched, sheared, etc., 
again subject to the constraint that they are destroyed and replaced by 
new ones wherever a spot of weak field appears. 
We do not currently have a more detailed mechanistic scenario of 
how exactly the folded structure with $\lpar/\lperp\sim\Beq/B_2$ is established.
It may be feasible to test these ideas numerically 
by solving MHD equations with viscosity locally determined by the magnetic-field 
strength according to \eq{Reeff_eq}. 

Let us assume that cluster fields do indeed have a folded structure 
with a direction reversal scale $\lperp\sim (0.01...0.1)L$, possibly 
determined by the argument given above.\footnote{Just 
such a structure appears to be evidenced by the polarised emission map 
from a radio relic in cluster A2256.\cite{Clarke_Ensslin}} 
The magnetic-energy spectrum then peaks at $k\sim1/\lperp$. 
What is the structure of the turbulence above and below 
this scale? At scales $l\ll\lperp$, the magnetic field reversing 
at the scale $\lperp$ will appear uniform and, in accordance with 
the old idea of Kraichnan\cite{Kraichnan} could support a cascade 
of Alfv\'en waves. This cascade can rigorously be shown to be described 
by the equations of Reduced MHD at collisionless scales all the way down 
to the ion gyroscale.\cite{SCDHHQ_gk} The currently accepted theory 
of such a cascade, primarily associated with the names of Goldreich and 
Sridhar,\cite{GS95} is based on the conjecture that at each scale, 
the Alfv\'en frequency is equal to the turbulent decorrelation rate. 
The result is a $k^{-5/3}$ spectrum of Alfv\'enic fluctuations 
--- this possibly explains what appears to be a $k^{-5/3}$ 
tail in the observed spectrum of magnetic energy for 
the Hydra A core.\cite{Vogt_Ensslin2} 
We cannot embark on a detailed discussion of the 
theory of the Alfv\'en-wave cascade here, so the reader 
is referred to Ref.~\onlinecite{SC_mhdbook} for a review and 
to Ref.~\onlinecite{SCDHHQ_gk} for the theoretical basis of 
extending this theory to collisionless scales. 

Above the reversal scale, $l\gg\lperp$, the cluster turbulence should 
resemble the saturated state of isotropic MHD turbulence: a magnetic-energy 
spectrum with a positive spectral index corresponding to folded fields 
and a kinetic-energy spectrum populated in the inertial range by 
a peculiar type of Alfv\'en waves that 
propagate along the folds (i.e., simultaneously perturbing the 
antiparallel magnetic field lines).\cite{SCHMM_ssim,SCTMM_stokes} 
This type of turbulence is also reviewed in Ref.~\onlinecite{SC_mhdbook}. 
It is probably of limited relevance for clusters 
because the Reynolds number in the ICM is not large enough to allow 
a well-developed inertial range. 

\Figref{fig_turb} summarizes the --- admittedly, rather speculative --- 
picture of cluster turbulence proposed above. 
We offer this sketch in lieu of conclusions. 
While we believe that the set of physical arguments that has led 
to it is not without merit, it is clear that much analytical, 
numerical and observational work is needed before a conclusion 
can truly be reached in the study of turbulence, magnetic fields 
and plasma physics in clusters of galaxies.

\begin{acknowledgments} 
Helpful discussions with T.~En{\ss}lin, G.~Hammett, R.~Kulsrud, E.~Quataert, 
and P.~Sharma are gratefully acknowledged. 
This work was supported by a UKAFF Fellowship, 
a PPARC Advanced Fellowship, King's College, Cambridge (A.A.S.) 
and by the DOE Center for Multiscale Plasma Dynamics. 
A.A.S.~also thanks the Royal Society (UK) and 
the APS for travel support. 
\end{acknowledgments}

\bibliography{sc_PoP}

\end{document}